\documentstyle[12pt]{article}
\begin{document}
\pagestyle{empty}
\vskip .7cm
\begin{center}
{\bf {BRST COHOMOLOGY AND HODGE DECOMPOSITION THEOREM \\
IN ABELIAN GAUGE THEORY}}

\vskip 2cm

{ R. P. MALIK} 
\footnote{ E-mail address: malik@boson.bose.res.in }\\
{\it S. N. Bose National Centre for Basic Sciences,}\\
{\it Block-JD, Sector-III, Salt Lake, Calcutta - 700 091, India}\\

\vskip .7cm

{\bf Abstract}

\end{center}

We discuss the Becchi-Rouet-Stora-Tyutin (BRST) cohomology and Hodge 
decomposition theorem for the two dimensional free $U(1)$ gauge theory. 
In addition to the usual BRST charge, we derive a local, conserved and 
nilpotent co(dual)-BRST charge under which the 
gauge-fixing term remains invariant. We express the Hodge decomposition
theorem in terms of these charges and the Laplacian operator.  We take 
a single photon state in the quantum Hilbert space and demonstrate the 
notion of gauge invariance, no-(anti)ghost theorem, transversality of 
photon and establish the topological nature of this theory by exploiting 
the concepts of BRST cohomology and  Hodge decomposition theorem. In fact, 
the topological nature of this theory is encoded in the vanishing 
of the Laplacian operator when equations of motion are exploited. On the 
two dimensional compact manifold, we derive two sets of topological 
invariants with respect to  the conserved and nilpotent 
BRST- and co-BRST charges and express the Lagrangian density of the
theory as the sum of terms that are BRST- and co-BRST invariants.
Mathematically, this theory captures together some of the key features
of both Witten- and Schwarz type of topological field theories.
\baselineskip=16pt

\vskip 1cm

\newpage

\pagestyle{headings}

\noindent 
{\bf 1 Introduction}\\

\noindent
The principles of local gauge invariance have played a very significant
role in the development of modern theoretical high energy physics up
to the energy scale of the order of grand unification. One of the
key features of these theories is the existence of first-class constraints
on them. The most natural and handy framework for the covariant canonical
quantization of such a class of theories is the BRST formalism [1,2]. In this 
scheme, we enlarge the phase space of the original gauge system
by incorporating the gauge-fixing and the Faddeev-Popov ghost terms in
the starting Lagrangian density. The ensuing theory is unitary and turns 
out to be invariant under a new, supersymmetric type and nilpotent BRST  
symmetry which incorporates the local gauge symmetry of the original Lagrangian
density in a subtle way. This symmetry is generated by a conserved and
nilpotent  BRST charge $Q_{B}$. The foot-prints of the original gauge theory
are encoded in the BRST charge because the requirement of the physical
state condition $Q_{B} \;| phys >\; = \;0 $ leads to the annihilation
of the physical states in the quantum Hilbert space by the 
first-class constraints
of the original gauge theory (see, $e.g.$, equation (5.3) below)
[3-6] \footnote{Some attempts have also been made to discuss the second class
constraints in the framework of BRST formalism (see, $e.g.$, Refs. [7,8]
and references therein).}.
The nilpotency of the  BRST charge $(Q_{B}^2 = 0)$ and the physical state
condition $(Q_{B}\;|phys> = 0)$ are the two key properties which are intimately
connected with the differential geometry and its application to cohomology
[5,6,9-12]. For instance, two  BRST closed states $(Q_{B}\; |phys> = 0,\;
Q_{B}\; |phys>^{\prime} = 0)$ in the quantum Hilbert space are said to be 
cohomologically equivalent if they differ by a  BRST exact state 
($i.e.,  |phys>^{\prime} \;=
|phys >\; + \;Q_{B} \;|\chi >$ for any nonzero $|\chi>$ in the Hilbert 
space). This property is analogous to the property of the exterior 
derivative $d (d^2 = 0)$  of differential geometry where two closed forms
($e.g.,  d f = 0, d f^{\prime} = 0$), defined on  a compact manifold, are 
cohomologically
equivalent if they differ by an exact form ($i.e., f^{\prime} = f + d g $).
One of the key theorems in the de Rham cohomology is the celebrated 
Hodge decomposition theorem defined on a compact manifold. This theorem
states that on this manifold any $p$-form $f_{p}$ can be decomposed into a
harmonic form $\omega_{p}$ $(\Delta \omega_{p} = 0, d \omega_{p} = 0, 
\delta \omega_{p} = 0)$ an exact form $ d g_{p-1}$ and a co-exact form 
$\delta h_{p+1}$ as follows:
$$
\begin{array}{lcl}
f_{p} = \omega_{p} + d g_{p-1} + \delta h_{p+1},
\end{array}\eqno(1.1)
$$
where $\delta (= \pm {*} d {*}; \delta^2 = 0)$ is the Hodge dual of $d$ 
and the Laplacian $\Delta$ is defined as 
$ \Delta = (d + \delta)^2 = d \delta + \delta d $ [9-12]. So far, 
the analogue of $d$ has been found out
as the conserved and nilpotent BRST charge $Q_{B}$ which generates a nilpotent
BRST symmetry for a locally gauge invariant Lagrangian density in any arbitrary
dimension of spacetime. It will be, therefore, an interesting endeavour
to express $\delta$ and $\Delta$ in terms of the {\it local conserved charges}
corresponding to some specific symmetry properties of a given 
BRST invariant Lagrangian density in any particular dimension of spacetime.

The purpose of the present paper is to shed some light on the analogues of
$\delta$ and $\Delta$ in the language of the nilpotent (for $\delta$), local, 
covariant and continuous symmetry properties of the free $U(1)$ gauge 
theory described by a BRST invariant Lagrangian density in two $ (1 + 1) $ 
dimensions(2D) of spacetime. Some attempts have been made towards this 
goal in any arbitrary dimension of spacetime for
the Abelian as well as non-Abelian 
gauge theories. However, the symmetry transformations turn out to be nonlocal
and noncovariant [13-16]. In the covariant formulation, the 
corresponding symmetries become even non-nilpotent and the nilpotency is 
restored only when certain specific restrictions are imposed [17].
We will demonstrate that for the 2D BRST invariant $U(1)$ gauge theory, 
a local, conserved and nilpotent 
(co)dual BRST charge $Q_{D}$ can be defined that generates a {\it new}
local, covariant and nilpotent
symmetry transformation under which  the gauge-fixing term 
$ \delta A = (\partial \cdot A) $ \footnote { Here the 
vector potential $A_{\mu}$ is defined through the one-form $ A =
A_{\mu}\; dx^{\mu}$. Furthermore, it can be easily seen that
the gauge-fixing term $ (\partial \cdot A)= \delta\; A$
is the Hodge dual of the two form $F = d A$ for the Abelian gauge theory
in any arbitrary dimension of spacetime
(see, $e.g.$, Ref. [11] for details).} remains invariant. This property
should be compared and contrasted with the usual BRST transformation 
under which the two-form $ F = d A $ remains invariant in the $U(1)$ 
gauge theory. Further, we show that the anticommutator of both these charges
$ W = \{ Q_{B}, Q_{D} \}$ is the analogue of the Laplacian operator
$\Delta$ and it turns out to be the Casimir operator for the extended
BRST algebra. It is, however, the topological nature of the 2D free
$U(1)$ gauge theory that $ W \rightarrow 0$ when equations
of motion are exploited and all the fields are assumed to fall off rapidly
at $ x \rightarrow \pm \infty $. To be more specific, we demonstrate that,
for a single photon state in the quantum Hilbert space, the BRST- and 
co-BRST symmetries are good enough to gauge away both the degrees of
freedom of photon and the free 2D $U(1)$ gauge theory becomes
topological in nature (see, $e.g.$, Ref. [18]). In the 
framework of BRST cohomology
and Hodge decomposition theorem, the topological nature of this theory 
is encoded in the vanishing of the Laplacian operator when equations of
motion are exploited. In fact, the on-shell expression for 
the Laplacian operator ($W$) encompasses
the left-over degrees of freedom in the theory. We derive two sets of 
topological invariants on the 2D compact manifold w.r.t. BRST- and co-BRST 
charges and express the Lagrangian density  as well as  energy momentum
tensor as the sum of  BRST- and co-BRST invariant parts. These  properties, 
together with symmetry considerations, are essential to establish
the topological nature of the 2D free $U(1)$ gauge theory.

The material of our work is organized as follows. In Sec. 2, we set
up the notations and give the bare essentials of the 
BRST formalism for the $U(1)$ gauge theory in any arbitrary
dimension of spacetime. In Sec. 3, we discuss various kinds of 
dualities in two dimensional free $U(1)$ gauge theory and derive
expressions for the co-BRST charge and the Casimir operator.  
Sec. 4 is devoted to the derivation of the extended BRST algebra.
The constraints on the physical states of the total Hilbert space are
obtained in Sec. 5. We take a single photon state as the harmonic state
of the Hodge decomposition theorem and demonstrate the strength of BRST
cohomology in the analysis of transversality of photon, gauge invariance,
no-ghost theorem, etc., and give a concise proof of the topological 
nature of this theory. We derive two sets of topological invariants on 
the 2D compact manifold w.r.t. BRST- and co-BRST charges in Sec. 6.
Finally, we discuss our main results,  make some concluding remarks and
propose some speculative ideas for future investigations.\\

\noindent             
{\bf  2 Preliminary: BRST Formalism }\\

\noindent
Let us begin with the BRST invariant Lagrangian density (${\cal L}_{b}$)
for the D-dimensional $U(1)$ gauge theory in the Feynman gauge 
(see, $e.g.$, [4-6])

$$
\begin{array}{lcl}
{\cal L}_{b} = - \frac{1}{4} F^{\mu\nu} F_{\mu\nu} + B (\partial \cdot  A)
+ \frac{1}{2} B^2 + i \bar C \;\Box \;C,
\end{array} \eqno(2.1)
$$
where $ \Box = \partial_{\mu} \partial^{\mu} (\mu = 0,1,2......D-1) $,
$ (\partial \cdot A) = \partial_{\mu} A^{\mu}, F_{\mu\nu} = \partial_{\mu}
A_{\nu} - \partial_{\nu} A_{\mu}$, $B$ is the Nakanishi-Lautrup auxiliary 
field and $\bar C (C)$ are (anti)ghost
fields with $ \bar C^2 = C^2 = 0$. The local gauge symmetry of the starting
Maxwell Lagrangian density ($- \frac{1}{4} F^{\mu\nu} F_{\mu\nu}$)
is now traded with the off-shell nilpotent $(\delta_{b}^2 = 0)$ 
BRST symmetry transformations 
$$
\begin{array}{lcl}
\delta_{b} A_{\mu} &=& \eta \;\partial_{\mu} \;C,\;\; \qquad
\;\;\;\delta_{b} F_{\mu\nu} = 0, \;\;\qquad \;\;\; \delta_{b} C = 0, \nonumber\\
\delta_{b} \bar C &=& i \;\eta\; B, \qquad \delta_{b} B = 0,
\qquad \delta_{b}(\partial \cdot  A)= \eta \;\Box \;C,  
\end{array} \eqno(2.2)
$$
where $\eta$ is an anticommuting ($ \eta C = - C \eta, \eta \bar C = -
\bar C \eta$) spacetime independent 
transformation parameter. The  following Lagrangian density obtained 
from (2.1) (with $ B = - (\partial \cdot A)$)
$$
\begin{array}{lcl}
{\cal L}_{B} = - \frac{1}{4} F^{\mu\nu} F_{\mu\nu}
- \frac{1}{2} (\partial \cdot A)^2 + i \bar C \Box C,
\end{array} \eqno(2.3)
$$
remains invariant under the on-shell ($\Box C = 0$) nilpotent 
($\delta_{B}^2 = 0$) BRST transformations
$$
\begin{array}{lcl}
\delta_{B} A_{\mu} &=& \eta \partial_{\mu} C, \qquad \delta_{B} C = 0,
\quad \delta_{B} F_{\mu\nu} = 0, \nonumber\\
\delta_{B} \bar C &=& - i \eta (\partial \cdot A), \qquad
\delta_{B} (\partial \cdot A) = \eta \Box C.
\end{array} \eqno(2.4)
$$
The conserved $( \dot Q_{b,B}= \partial_{0} Q_{b,B} = 0)$ and nilpotent
$(Q_{b,B}^2 = 0)$ BRST charge $(Q_{b,B})$

$$
\begin{array}{lcl}
Q_{b,B} = {\displaystyle \int d^{(D-1)} x}\; \bigl ( B \dot C - \dot B C \bigr )
\equiv {\displaystyle \int d^{D-1} x}\; 
\bigl [\; \partial_{0} (\partial \cdot A) C - 
(\partial \cdot A) \dot C\; \bigr ],
\end{array} \eqno(2.5)
$$
is the generator of the transformations (2.2) and (2.4) as the following
equation
 
$$
\begin{array}{lcl}
\delta_{k} \Phi = - i \;\eta \;\bigl [\; \Phi, Q_{k} \;
\bigr ]_{\pm}, \qquad k = b, B,
\end{array} \eqno(2.6)
$$
where $ (+)- $ stands for the (anti)commutator 
(depending on whether the generic field $\Phi$ is (fermionic)bosonic), 
generates the above transformations if one exploits the covariant canonical 
(BRST) quantization of the Lagrangian density (2.3)

$$
\begin{array}{lcl}
&& [ A_{0} (x,t), (\partial \cdot A)(y,t) ] = - i \delta^{(D-1)} (x - y),
\nonumber\\
&& [ A_{i} (x,t), E_{j} (y,t) ]\; =\; i\; \delta_{ij}\; 
\delta^{(D-1)} (x - y),\nonumber\\
&& \{ C (x,t), \dot {\bar C} (y,t) \}\; =\; \delta^{(D-1)} (x - y),\nonumber\\
&& \{ \bar C (x,t), \dot C (y,t) \} \;= \;- \delta^{(D-1)} (x - y),
\end{array}\eqno(2.7)
$$  
and all the rest of the (anti)commutators are zero.

The invariance of the ghost action $ I_{F.P.}
= i \int d^{D} x \; \bar C\; \Box\; C $ 
under
the global scale transformations : $ C \rightarrow  e^{\lambda} C, \bar C
\rightarrow  e^{- \lambda} \bar C $ leads to the derivation of a conserved
charge ($Q_{g}$)
$$
\begin{array}{lcl}
Q_{g} = - i {\displaystyle \int d^{(D-1)} x }\;
\bigl (\; C\; \dot {\bar C} + \bar C\; \dot C \;\bigr ). 
\end{array}\eqno(2.8)
$$
Furthermore, the discrete symmetry:
$ C \rightarrow \pm i \bar C, \bar C \rightarrow  \pm i C $ invariance
of $I_{F.P.}$ leads to
the existence of a nilpotent and conserved anti-BRST charge $Q_{AB}$
whose expression as well as the symmetry transformations it generates, can
be obtained from equations (2.5), (2.2) and (2.4) by the substitution:
$ C \rightarrow  \pm i \bar C$. The following BRST algebra  
$$
\begin{array}{lcl}
&& \{ Q_{B}, Q_{B} \}\;\;\; =\;\;\; \{ Q_{AB}, Q_{AB} \}\;\; \;= \;\;0, \nonumber\\
&& \{ Q_{B},   Q_{AB} \} = Q_{B} Q_{AB}\; +\; Q_{AB} Q_{B} = 0,\nonumber\\
&& i [  Q_{g}, Q_{B} ] = + Q_{B}, \quad 
i [  Q_{g}, Q_{AB} ] = - Q_{AB}, 
\end{array}\eqno(2.9)
$$
states that the ghost number is $ +1$ for $Q_{B}$ and
$-1$ for $Q_{AB}$.

At this stage, it is important to pin-point some of the salient features
which will be relevant for our further discussions. First of all, the 
statements that have been made above, are valid in any 
arbitrary dimension of spacetime.
Secondly, it can be checked that the transformations generated by $Q_{B}$
and $Q_{AB}$ anticommute $( \delta_{B} \delta_{AB} + \delta_{AB} \delta_{B}
= 0)$ when they act on any field. Thirdly, under both the transformations,
it is the two-form $ F = d A $ (or $ F_{\mu\nu} = \partial_{\mu} A_{\nu}
- \partial_{\nu} A_{\mu}$) that remains invariant and {\it not} the gauge-
fixing term $ \delta\; A $ ( or $\partial \cdot A )$ which is dual to it.
Finally, it is obvious that the anti-BRST charge $Q_{AB}$ is not the analogue
of the dual exterior derivative $ \delta = \pm {*} d {*} $ in 
the discussion of the cohomological aspects of  BRST formalism. 
As a consequence, in any arbitrary spacetime dimension, 
there are no analogues of the dual exterior derivative
$\delta$ and the Laplacian $ \Delta $ in the language of the nilpotent, 
local, covariant and continuous symmetry properties of the BRST invariant
Lagrangian (2.3) or (2.1). Hence, the Hodge decomposition theorem defined 
on a compact manifold, can not be implemented in the quantum Hilbert space 
of such theories. In 2D of spacetime, however, we shall demonstrate that 
symmetries of the BRST invariant Lagrangian density are such
that there is one-to-one correspondence with the (dual)exterior derivative 
and the Laplacian operator of differential geometry and 
the (dual)BRST charge and the Casimir operator of the extended 
BRST algebra. \\

\noindent
{\bf 3 BRST-Type Symmetries and Dualities in 2D}\\

\noindent
In addition to the symmetries: $ C \rightarrow  \pm i \bar C,\; \bar C
\rightarrow  \pm i C $, the ghost action $i 
\int \; d^2 x\;\;\bar C \;\Box \;C $ in 2D 
has another symmetry; namely,
\footnote{ We adopt here the notations in which the 2D flat Minkowski
metric is: $\eta_{\mu\nu} = $ diag (+1, -1) and $ \Box = \eta^{\mu\nu}
\partial_{\mu} \partial_{\nu} = \partial_{0} \partial_{0} -
\partial_{1} \partial_{1}, F_{01} = 
\partial_{0} A_{1} - \partial_{1} A_{0} = E
= F^{10}, \varepsilon_{01} = \varepsilon^{10} = +1, 
(\partial \cdot A) = \partial_{0} A_{0} - \partial_{1} A_{1}$. }.
$$
\begin{array}{lcl}
\partial_{\mu} \rightarrow  \pm\; i\; \varepsilon_{\mu\nu}\; \partial^{\nu},
\qquad \;\;\varepsilon_{\mu\nu} \varepsilon^{\mu\lambda} 
= - \delta_{\nu}^{\lambda},
\end{array} \eqno (3.1)
$$
under which the D'Alembertian $ \Box $ remains invariant. As a consequence, 
an analogue of the symmetry (2.4) for the Lagrangian density (2.3), can be
obtained due to the additional symmetry property of the 2D ghost term.
These two symmetries are juxtaposed  as 

$$
\begin{array}{lcl}
&&\delta_{B} A_{\mu} = \eta \partial_{\mu} C, \;\;\;\;\qquad \;\;
\delta_{D} A_{\mu} = - \eta \varepsilon_{\mu\nu} \partial^{\nu} \bar C,
\nonumber\\
&& \delta_{B} C = 0, \;\;\;\;\;\;\; \qquad \;\; \;\;\;\;\;
\delta_{D} \bar C = 0, \nonumber\\
&& \delta_{B} \bar C = - i \eta (\partial \cdot A), \qquad
\delta_{D} C = - i \eta E, \nonumber\\
&& \delta_{B} E = 0, \;\;\;\;\;\;\qquad \;\;\; \;\;
\;\;\;\;\delta_{D} (\partial \cdot A) = 0,
\nonumber\\
&& \delta_{B} (\partial \cdot A) = \eta \Box C, \qquad \;\;
\delta_{D} E = \eta \Box \bar C, 
\end{array}\eqno(3.2)
$$
where we have taken : $ C \rightarrow  + i \bar C, \partial_{\mu}
\rightarrow  + i \varepsilon_{\mu\nu} \partial^{\nu} $ in deriving
symmetry transformations $\delta_{D}$ from the BRST symmetries 
$\delta_{B}$
\footnote{ Here and in what follows, we shall take only the ($+$) sign
in the transformations: $ C \rightarrow  \pm i \bar C, \bar C
\rightarrow  \pm i C, \partial_{\mu} \rightarrow  \pm i 
\varepsilon_{\mu\nu} \partial^{\nu} $. However, analogous
statements will be valid if we take ($-$) sign.}.  
It can be checked that both these symmetry transformations
are on-shell nilpotent for the Lagrangian (2.3). Furthermore, under the 
above transformations, it can be checked that the 2D BRST invariant
Lagrangian density
$$
\begin{array}{lcl}
{\cal L}_{B} = \frac{1}{2} E^2 - \frac{1}{2} (\partial \cdot A)^2
+ i \;\bar C\; \Box\; C,
\end{array}\eqno(3.3)
$$
transforms to itself modulo some total derivative terms. We christen
the $\delta_{D}$ transformations in (3.2) as the dual-BRST transformations
because in contrast to $\delta_{B}$ transformations where electric field 
$E$ is invariant, in the case of $\delta_{D}$, it is the gauge-fixing
term $(\partial \cdot A)$ that remains invariant \footnote{ As per our 
definition in the introduction, the gauge-fixing term $\delta A = (\partial
\cdot A) $ with $ \delta = \pm {*} d {*} $ is the Hodge dual of the two-form 
$ F = d A$ which is the electric field $E$ here in 2D.}. Moreover, it
is interesting to see that under $ \partial_{\mu} \rightarrow   \pm
i \varepsilon_{\mu\nu} \partial^{\nu}$, the gauge-fixing term and 
the kinetic energy
term in the Lagrangian density (3.3), transform to each other.
This mutual exchange can be obtained even with the transformation $ A_{\mu}
\rightarrow \pm i \varepsilon_{\mu\nu} A^{\nu}$. However, the latter
transformation is not the symmetry of the 2D ghost action with which the 
BRST-type symmetries are connected. Thus, we shall call the duality 
transformations as the ones in which $ C \rightarrow \pm i \bar C, \bar C
\rightarrow  \pm
i C, \partial_{\mu} \rightarrow  \pm i \varepsilon_{\mu\nu} \partial^{\nu}$.
This will be the analogue of the Hodge dual operation ${*}$ in the
differential geometry connected with the cohomological aspects of differential
forms defined on a compact manifold.

In 2D, the field strength tensor  $F_{\mu\nu}$ has only one independent
component which corresponds to the electric field $E$ and there is no magnetic
field $B$. Thus, the usual duality invariance of the Maxwell equations
under $ E_{i} \rightarrow  B_{i}, B_{i} \rightarrow  - E_{i}$ in 4D cannot
be obtained here in 2D. However, the gauge-fixing term $ \partial \cdot A 
= \partial_{\mu} A^{\mu} $
and the electric field $ E = - \varepsilon_{\mu\nu} \partial^{\mu} A^{\nu} $
are like scalar and pseudoscalar in 2D. Thus, a duality between gauge-fixing
and the electric field can be defined. For instance, it can be seen that the
following Maxwell equations
$$
\begin{array}{lcl}
\partial_{\mu} F^{\mu\nu} + \partial^{\nu} (\partial \cdot A) = 0,
\end{array}\eqno(3.4)
$$
lead to two independent equations 
$$
\begin{array}{lcl}
\partial_{0} E + \partial_{1} ( \partial \cdot A) &=& 0, \nonumber\\
\partial_{1} E + \partial_{0} ( \partial \cdot A) &=& 0,
\end{array}\eqno(3.5)
$$
which remain invariant under the exchange of gauge field $E$  and gauge-fixing
term $(\partial \cdot A)$. The same symmetry can be achieved if the
derivatives $\partial_{0}$ and $\partial_{1}$ are exchanged with each-other.
Furthermore, if we define $ \tilde E = E + i (\partial \cdot A), \; 
\tilde B =
E - i (\partial \cdot A) $, the equation (3.5) can be re-expressed as:
$ \partial_{0} \tilde E + i \partial_{1} \tilde B = 0;\; \partial_{0} \tilde B
- i \partial_{1} \tilde E = 0 $, which respects duality-type invariance:
$ \tilde E \rightarrow \tilde B, \tilde B \rightarrow - \tilde E $.
Now we take the Hodge decomposition of the one-form $A$ in terms of the
analogues of a scalar field $\phi$  and a pseudoscalar field $\lambda$ in 2D
$$
\begin{array}{lcl}
A_{\mu}\; dx^{\mu} = \partial_{\mu} \phi \;dx^{\mu} 
+ \varepsilon_{\mu\nu}\; \partial^{\nu} \lambda \;d x^{\mu},
\end{array}\eqno(3.6)
$$
and define $ \kappa = \phi + i \lambda$ and $ \xi = \phi - i \lambda $, the
Maxwell equations (3.5) can be recast into
$$
\begin{array}{lcl}
\Box \; (\partial_{0} \kappa -\; i \partial_{1} \xi)   &=& 0, \nonumber\\
\Box \;(\partial_{0} \xi + \;i\; \partial_{1} \kappa)  &=& 0,
\end{array}\eqno(3.7)
$$
which have $ \kappa \rightarrow  \xi, \;\; \xi \rightarrow - \kappa $
symmetry\footnote{ Notice that eventhough we denote $\kappa, \xi$ and 
$\tilde E, \tilde B$ by different symbols, they are complex conjugate
to each-other in each pair because $ E, (\partial \cdot A), \phi$ and 
$\lambda$ are taken to be real.} that resembles the usual duality symmetry 
of the Maxwell equations in more than 2D. In fact, due to the decomposition 
(3.6), the Lagrangian density (2.3) can be rewritten in the following form
 
$$
\begin{array}{lcl}
{\cal L}_{B} = \frac{1}{2} ( \Box \lambda )^2 - \frac{1}{2} ( \Box \phi )^2
+ i \bar C\; \Box \; C,
\end{array}\eqno(3.8)
$$
which respects the symmetry transformations $ \partial_{\mu} \rightarrow
\pm i \varepsilon_{\mu\nu} \partial^{\nu}, C \rightarrow \pm i \bar C,
\bar C \rightarrow  \pm i C, \lambda \rightarrow  \pm i \phi, \phi
\rightarrow  \pm i \lambda $. For this Lagrangian density, the symmetry
transformations $\delta_{B}$ and $\delta_{D}$ are

$$
\begin{array}{lcl}
&& \delta_{B} \phi = \eta \; C, \;\;\;\qquad \;\;\;
\delta_{D} \phi = 0, \nonumber\\
&& \delta_{B} \lambda = 0,\; \;\;\;\;\;\qquad \;\;\;\; 
\delta_{D} \lambda = - \eta \bar C,\nonumber\\
&& \delta_{B} C = 0, \;\;\;\;\qquad \;\;\;\;\;
\delta_{D} C = i \eta \Box \lambda, \nonumber\\
&&\delta_{B} \bar C = - i \eta \Box \phi, \qquad \;
\delta_{D} \bar C = 0.
\end{array}\eqno(3.9)
$$
As symmetry transformations (3.2) and (3.9) are fermionic in nature, it
is straightforward to check that their anticommutator $ \{ \delta_{B},
\delta_{D} \} $ would also lead to a symmetry transformation for the
Lagrangian (2.3). Such a symmetry transformation $ \delta_{W} 
= \{ \delta_{B} \delta_{D} \} $  with the transformation parameter
$ \zeta = - i\; \eta \;\eta^{\prime}$  is

$$
\begin{array}{lcl}
\delta_{W} C\; &=& \;\;\;\delta_{W}\; \bar C\;\;\;\; = \;\;0, \nonumber\\
\delta_{W} A_{0} &=& \zeta\; \bigl ( \partial_{0} E + \partial_{1} 
(\partial \cdot A) \bigr ), \nonumber\\
\delta_{W} A_{1} &=& \zeta \;\bigl ( \partial_{1} E + \partial_{0}
(\partial \cdot A),  
\end{array}\eqno(3.10)
$$
where $\eta$ and $\eta^{\prime}$ are the anticommuting transformation
parameters for the transformations $\delta_{B}$ and $\delta_{D}$. 
For the transformations $\delta_{W}$, it is the ghost
term of the Lagrangian density (2.3) which remains invariant and the
gauge-fixing as well as the electric fields transform. These transformations
can be recast in terms of $\phi$ and $\lambda$ fields of the Lagrangian density
(3.8) as
$$
\begin{array}{lcl}
\delta_{W} \phi &=& - \zeta \; \Box \lambda, \qquad \delta_{W}  C = 0,
\nonumber\\
\delta_{W} \lambda &=& - \zeta\; \Box \phi, \qquad
\delta_{W} \bar C = 0.
\end{array}\eqno(3.11)
$$
Under all the above three symmetries, the action
$ S = \int \;d^2 x\; {\cal L}_{B} $ remains invariant because $ \delta_{k}
S = 0 $ for $ k = B, D, W $ if all the fields fall off rapidly at
$ x \rightarrow \pm \infty $.\\

\noindent
{\bf 4 Extended BRST Algebra and Conserved Quantities}\\

\noindent
The additional two local, continuous and covariant symmetries $\delta_{D}$ 
(with $ \delta_{D}^2 = 0)$ and $\delta_{W}$ of the
BRST invariant Lagrangian density for the $U(1)$ gauge theory in 2D
lead to the following conserved charges due to the Noether theorem
$$
\begin{array}{lcl}
Q_{D} = \int \; dx \Bigl [\; E \;\dot {\bar C} - \dot E\; \bar C \;\Bigr ],
\qquad W = \int\;dx  \Bigl [\; \partial_{0} (\partial \cdot A) E
- (\partial \cdot A)\; \dot E \; \Bigr ],
\end{array}\eqno(4.1)
$$
which are in addition to the three charges $ Q_{B}, Q_{AB} $ and
$Q_{g}$  of Sec. 2 (cf. equations (2.5) and (2.8)). However, the
presence of the symmetry $ C \rightarrow \pm i \bar C $ and
$ \bar C \rightarrow  \pm i C$, leads to the existence of the analogues
of transformations (3.2), (3.9--3.11) under which the Lagrangian density
(2.3) remains invariant. The set of all these charges for 2D case is
$$
\begin{array}{lcl}
Q_{B} &=& {\displaystyle \int dx }\; 
\Bigl [\;\partial_{0} (\partial \cdot A) C
- (\partial \cdot A) \; \dot C \;\Bigr ], \qquad \;\;
Q_{AB} = i {\displaystyle \int  dx} \;
\Bigl [\; \partial_{0} (\partial \cdot A)
\bar C - (\partial \cdot A) \dot {\bar C}\; \Bigr ], \nonumber\\
Q_{D} &=& {\displaystyle \int dx} \;\Bigl 
[\; E \;\dot {\bar C} - \dot E\; \bar C \;\Bigr ],
\;\;\;\;\;\;\;\;\;\;\qquad\;\;\;\;\;\;\;\;\; Q_{AD} = i 
{\displaystyle \int dx} \; 
\Bigl [\; E \;\dot C - \dot E\; C\; \Bigr],
\nonumber\\
W &=& {\displaystyle \int dx} \;
\Bigl [\; \partial_{0} ( \partial \cdot A) E -
\dot E ( \partial \cdot A)\; \Bigr ],\;\;\; \qquad \;\;\;
Q_{g} = - i {\displaystyle \int dx} \;
\Bigl [\; C\; \dot {\bar C} + \bar C\; \dot C \;\Bigr ]. 
\end{array}\eqno(4.2)
$$
If we exploit the covariant canonical (anti)commutators of equation (2.7),
these conserved charges obey the following algebra
$$
\begin{array}{lcl}
&& [ W, Q_{k}] = 0, k = B, D, AB, AD, g, \quad \nonumber\\
&&Q_{B}^2\; = \;Q_{AB}^2\; = \;Q_{D}^2\; = \;Q_{AD}^2\; = 0,\nonumber\\
&& \{ Q_{B}, Q_{D} \}\; = \;\{ Q_{AB}, Q_{AD} \}\; =\; W, \nonumber\\
&& i [ Q_{g}, Q_{B} ] = + Q_{B}, \;\;\;\quad\;\;
i [ Q_{g}, Q_{AB} ] = - Q_{AB}, \nonumber\\
&& i [ Q_{g}, Q_{D} ] = - Q_{D}, \qquad \;
i [ Q_{g}, Q_{AD} ] = + Q_{AD},
\end{array} \eqno(4.3)
$$
and all the rest of the (anti)commutators turn out to be zero.
A few remarks are in order. First of all, we see that the operator
$W$ is the Casimir operator for the whole algebra and its ghost
number is zero. The ghost number of $Q_{B}$ and $Q_{AD}$ is $+1$ and that 
of $Q_{D}$ and $Q_{AB}$ is $-1$. There exist four nilpotent and conserved
charges which generate (anti)BRST and (anti)dual BRST transformations.
Now given a state $ |\psi>$ in the quantum Hilbert space with ghost 
number $n$ ($ i.e., i Q_{g} | \psi> = n |\psi >$), it is straightforward
to see that:
$$ 
\begin{array}{lcl}
i Q_{g}\; Q_{B} |\psi> &=& ( n + 1)\; Q_{B} |\psi>, \nonumber\\
i Q_{g}\; Q_{D} |\psi> &=& ( n - 1)\; Q_{D} |\psi>, \nonumber\\
i Q_{g}\; W \;    |\psi> &=& n\; W \;|\psi>,
\end{array}\eqno(4.4)  
$$ 
which demonstrate that, whereas $W$ keeps the ghost number of a state intact
and unaltered, the operator $Q_{B}$ increases the ghost number
by one and $Q_{D}$ reduces this number by one. This property is similar
to the operation of a Laplacian, an exterior derivative and a dual exterior
derivative on a $n$-form defined on a compact manifold. Thus,
we see that the degree of the differential form is analogous to the ghost
number in the Hilbert space, the differential form itself is analogous to the
quantum state in the Hilbert space, a compact manifold has an analogy with
the quantum Hilbert space and $d$, $\delta$ and $ \Delta = d \delta
+ \delta d $ are $Q_{B}$ , $Q_{D}$ and $W$ respectively. It is a notable
point that $d$ and $\delta$  can be also identified with
$Q_{AB}$ and  $Q_{AD}$ because of the fact that the gauge parameter
of the original local gauge symmetry can be replaced either by a
ghost field or by an antighost field due to the symmetry properties
of the ghost action.

There are other conserved quantities in the theory due to the form
of the equations of motion in 2D $U(1)$ gauge theory. For instance,
it can be seen, from the equation of motion (3.5), that the 
gauge-fixing term $(\partial \cdot A)$ and the electric field $E$ are
conserved quantities.  Since these quantities are conjugate momenta
w.r.t. $A_{0}$ and $A_{1}$ fields, they commute with each
other. In fact, one can construct infinite number of commuting
conserved quantities with $E$ and $(\partial \cdot A)$ because
of the fact that the linear equation (3.5) is integrable (see, $e.g.,$ 
Ref. [19] for the nonlinear PDE like Boussinesq). 
Some of these quantities are:

$$
\begin{array}{lcl}
&&I_{0} = \int \; dx \;E, \;\;\qquad I_{1} = \int \; dx \;(\partial \cdot A),
\qquad\;\;\; I_{2} = \int \; dx\; E (\partial \cdot A),\nonumber\\
&&I_{3} = \frac{1}{2}\; \int \; dx \;\bigl [ E^2 + (\partial \cdot A)^2 
\bigr ], \qquad \;I_{4} = \int \; dx \;\bigl [ \partial_{0} 
(\partial \cdot A)\; E -
(\partial \cdot A) \dot E \bigr ],\nonumber\\ 
&&I_{k} = \;\int dx \;\;Z_{k},
\;\;\;\;\;\;\;\;\; \qquad \;
\;\;\;\;\;\;\;\;\;\;\;\;k = 5, 6, 7, 8,\nonumber\\
&&I_{9}= \frac{1}{2}\int\; dx \; 
\bigl [\; Z_{5}^2\; + \;Z_{6}^2 \;\bigr ], \;\;\;\qquad\;\;\;
I_{10} = \frac{1}{2}\int \; dx \bigl [\; Z_{7}^2\; 
+\; Z_{8}^2 \;\bigr ],\nonumber\\
&&....................................................\nonumber\\
&&....................................................
\end{array} \eqno(4.5)
$$
where the quantities $Z_{k}$ are functions of $(\partial \cdot A)$
and $E$ given by:
$$
\begin{array}{lcl}
Z_{5} &=& Cos (\partial \cdot A)\; Cos (E), \qquad \;\;\;
Z_{6} = Sin (\partial \cdot A)\; Sin (E), \nonumber\\
Z_{7} &=& Cos (\partial \cdot A) \; Sin (E), \qquad \;\;\;
Z_{8} = Sin (\partial \cdot A)\; Cos (E).
\end{array}\eqno(4.6)
$$
It can be checked that the $Z_{k}'s$ obey the following equations
$$
\begin{array}{lcl}
{\displaystyle \frac{\partial Z_{k} }{\partial t}} \;  &=& 
{\displaystyle \frac{\partial Z_{k}^{\prime}}
{\partial x}}, \qquad \;\;\;\; \Box \;Z_{k} \;=\; 0, 
\quad (k = 5, 6, 7, 8), \nonumber\\
Z_{5}^{\prime} &=& Z_{6}, \quad Z_{6}^{\prime} = Z_{5},
\quad Z_{7}^{\prime} = - Z_{8}, \quad Z_{8}^{\prime} = - Z_{7},
\end{array}\eqno(4.7)
$$
if we assume the validity of the on-shell conditions (3.5) for the
$U(1)$ gauge fields $A_{\mu}$. Thus, it will be noticed that even 
without taking recourse to the (anti)ghost
fields, the conserved quantity $I_{4}$, which is the Casimir operator $W$
of the algebra (4.3), can be constructed just by looking at the equations
of motion for the $U(1)$ gauge field $A_{\mu}$. In the algebra (4.3),
$W$ emerges due to the anticommutation relation between two charges which
contain (anti)ghost fields. It is not clear whether all the other
conserved charges in (4.5) can be expressed as the anticommutator of
two fermionic charges which contain (anti)ghosts.

The equations of motion in the ghost sector $ \Box C = \Box \bar C = 0$
are such that the following quantities (as functions of (anti)ghost fields
and their conjugate momenta) 
$$
\begin{array}{lcl}
G_{0} &=& -i \int\; dx \;\bigl ( C \;\dot {\bar C} + \bar C\; \dot C \bigr ),
\nonumber\\
G_{1} &=& - i \int\; dx \;\dot C (x,t), \quad 
G_{-1} = i \int\; dx \;\dot {\bar C} (x,t),
\end{array}\eqno(4.8)
$$
are conserved on-shell if we assume that all the fields fall off
rapidly at $ x \rightarrow  \pm \infty $ and there is no nontrivial
topology at the boundary of the manifold. It can be seen
that $G_{0}$ is the expression of the ghost charge $Q_{g}$. These charges 
obey the following algebra if we exploit the anticommutators of (2.7)
$$
\begin{array}{lcl}
&& G_{1}^2 = \frac{1}{2}\; \{ G_{1}, G_{1} \} =\; 0, \;\;\;\;\qquad \;\;\;
G_{-1}^2 = \frac{1}{2} \{ G_{-1}, G_{-1} \} =\; 0,
\nonumber\\
&& \{G_{1}, G_{-1} \} = 0, \quad 
i \{ G_{0}, G_{1}\} = + G_{1}, \quad i \{ G_{0}, G_{-1} \} = - G_{-1}.
\end{array}\eqno(4.9)
$$
The above algebra shows that the ghost number of $G_{\pm1}$ is $\pm 1$
and they generate trivial symmetry transformations for (anti)ghost fields
where these fields transform by a constant. 
Because of the fermionic nature of (anti)ghost fields, it appears that
no more conserved quantities can be constructed based on the equations
of motion $ \Box C = \Box \bar C = 0 $.

With the help of (4.7), one can construct other conserved charges which
are analogous to $Q_{B}$ and $Q_{D}$ and contain $Z_{k}'s$ and (anti)ghost
fields; namely,
$$
\begin{array}{lcl}
J_{s} &=& \int \;dx \;\bigl [\; (\partial_{0} \;Z_{s}) \; C - Z_{s}\; \dot C
\;\bigr ],\nonumber\\
J_{r}^{\prime} &=& \int \;dx \;\bigl [\; Z_{r}^{\prime} \;\dot {\bar C} - 
(\partial_{0} Z_{r}^{\prime}) \;  \bar C\; \bigr ],
\end{array}\eqno(4.10)
$$
where $ r, s = 5, 6, 7, 8 $  of (4.6) and (4.7).
It is clear that the anticommutator of $J_{s}$ and $J_{r}^{\prime}$ will also
produce analogue of $I_{4}$. All the conserved quantities in (4.7), (4.8)
and (4.10) generate certain symmetries. However, the conserved quantity
$W = I_{4}$ is singled out from all the rest of the conserved quantities
in (4.5) and (4.10) because of its very special nature. It turns out that, 
without resorting to the on-shell condition (3.5), the Lagrangian density (3.3)
transforms to a total derivative $ \partial_{\mu} [\; E \partial^{\mu} 
(\partial \cdot A) - \partial^{\mu} E \; (\partial \cdot A) ] $ under (3.10) 
which is generated by $I_{4} = W$. The  rest of the conserved quantities 
generate nontrivial symmetry of the Lagrangian density (3.3) only when the 
on-shell conditions (3.5) and $ \Box C = \Box \bar C = 0 $ are exploited. For 
instance, $I_{2}$ generates the transformations:
$ \delta_{2} C = \delta_{2} \bar C = 0,\; \delta_{2} A_{0} = - \rho\; E,\;
\delta_{2} A_{1} = \rho\; (\partial \cdot A)$ where $\rho$ is a constant
parameter. This transformation is a symmetry transformation of (3.3)
only if the on-shell equations (3.5) are utilized
\footnote{ Similar kind of symmetry has been pointed out in the context of
2D anomalous gauge theory where usual gauge symmetry turns out to be
the on-shell symmetry of the chiral Schwinger model [20].}. Thus, the
Casimir operator $ W = I_{4}$ is unique in some sense.\\

\noindent
{\bf 5 BRST Cohomology and Physical States}\\

\noindent
It is obvious from the algebra (4.3) and the consideration of the ghost number
of states ($Q_{B} |\psi>, Q_{D} |\psi> $ and $ W |\psi> $ in (4.4)) that 
one can now implement the Hodge decomposition theorem in the language of 
the BRST and dual-BRST charges
$$
\begin{array}{lcl}
|\psi >_{n} = |\omega >_{n} +\; Q_{B} |\; \theta >_{n-1} 
+\; Q_{D}\; | \chi >_{n+1},
\end{array}\eqno(5.1)
$$
by which, any state $|\psi >_{n}$ in the quantum Hilbert space with 
ghost number $n$ can be decomposed into a harmonic state $|\omega >_{n}$,
a BRST exact state $ Q_{B} |\theta >_{n-1} $ and a dual-BRST exact state
$ Q_{D} | \chi >_{n +1} $. To refine the BRST cohomology, however, we have
to choose a representative state from the
total states of (5.1) as a physical state. We take  here the physical state
as the harmonic state:  $| phys > = | \omega >$. By definition, 
such a state would satisfy the following conditions:
$$
\begin{array}{lcl}
 Q_{B}\; | phys > = 0 , \qquad Q_{D}\; | phys > = 0,
\qquad  W \;|phys > = 0,
\end{array} \eqno(5.2)
$$
which leads to the following constraints on the physical states:
$$
\begin{array}{lcl}
- \Pi^{0} =\; (\partial \cdot A)\;\; | phys > & = & 0,\nonumber\\
- \partial_{1} E =\; \partial_{0} (\partial \cdot A)\;\; |phys> &=& 0,
\nonumber\\
- \varepsilon_{\mu\nu} \partial^{\mu} A^{\nu} = \;E\;\; |phys > &=& 0,
\nonumber\\
- \partial_{1} (\partial \cdot A) = \;\dot E \;\;|phys > &=& 0.
\end{array}\eqno(5.3)
$$
At this juncture, it is worthwhile to point out that the latter
pair of constraints are related to the former ones by duality
transformations $ (\partial \cdot A) \rightarrow  \pm\; i\; E,\;
E \rightarrow  \pm\; i\; (\partial \cdot A) $. This demonstrates that
mathematically different looking theories in 2D, with different
constraints ($cf$. (5.3)), are same theories because they are
related to each-other by the duality transformations between 
gauge-fixing term $ (\partial \cdot A) $ and classical gauge 
field $E$. The first pair of constraints are obtained by
the requirement that $ Q_{B} | phys > = 0 $ which demonstrate
that the first class constraints $ \Pi^{0}\approx 0 $ (momentum
w.r.t. $A_{0}$ field)  and the Gauss law constraint 
($ \partial_{1} E \approx 0$) annihilate the physical state. The latter
constraints are also interesting as they lead to the proof of 
topological nature of 2D free $U(1)$ gauge theory (see, $e.g.$ Sect. 6).
In fact, they restrict the physical (harmonic) states to a sector
carrying zero electric flux. As a result,  even the electric field turns
out to be physically superfluous (in some sense)
because of the presence of the new co-BRST symmetry.
This happens due to the fact that
there are no propagating degrees of freedom in the theory. Furthermore,
the above restrictions imply the masslessness ($ \Box A_{\mu} = 0$) of
the photon {\it from the BRST- and co-BRST symmetries alone}. In the normal
formulation with a single charge ($Q_{B}$), the topological nature of the
2D photon is proven by transversality requirement ($ \partial \cdot A 
|phys> = 0$) which emerges from symmetry considerations
($ Q_{B} |phys> = 0$)  and masslessness
condition which emerges from equation of motion ($ \Box A_{\mu} = 0$).

We shall dwell a bit more on the constraints (5.3) in the
phase space representation and demonstrate that the physical
state conditions (5.2) contain a lot of information about
the gauge theory. Because of the simple form of the
equations of motion $ \Box A_{\mu} = 0, \Box C = 0 $ and
$ \Box \bar C = 0 $, it is very convenient to express
the fields $A_{\mu}, C $ and $ \bar C$ in terms of the
normal mode expansion [21]
$$
\begin{array}{lcl}
A_{\mu}( x, t )&=&  
{\displaystyle \int \frac{ d k} { (2\pi)^{1/2} (2 k^{0})^{1/2}}}
\Bigl [\; a_{\mu}(k) e^{- i k \cdot x } + a_{\mu}^{\dagger}(k) 
e^{ i k \cdot x}\; \Bigr ],
\nonumber\\
C (x,t) &=& 
{\displaystyle \int \frac{ d k } { (2 \pi)^{1/2} (2 k^{0})^{1/2}}}
\Bigl [\; c(k) e^{- i k \cdot x} + c^{\dagger} (k)
e^{i k \cdot x} \;\Bigr ], \nonumber\\
\bar C (x,t) &=& 
{\displaystyle \int \frac { d k } { (2\pi)^{1/2} (2 k^{0})^{1/2} }}
\Bigl [\; b(k) e^{ - i k \cdot x} + b^{\dagger} (k)
e^{i k \cdot x} \;\Bigr ], 
\end{array}\eqno(5.4)
$$
where $ k_{\mu} $ is the 2D momenta with the components 
$ (k_{0}, k = k_{1})$.  The symmetry transformations (3.2)
can now be exploited to obtain the (anti)commutation relations.
These are:
$$
\begin{array}{lcl}
&& [ Q_{B}, a^{\dagger}_{\mu}(k) ] = - k_{\mu} c^{\dagger} (k),
\qquad\;\;\; 
[ Q_{D}, a^{\dagger}_{\mu}(k) ] = \varepsilon_{\mu\nu} k^{\nu}
b^{\dagger} (k), \nonumber\\
&& [ Q_{B}, a_{\mu} (k) ] = k_{\mu} c(k),\;\; \;\;\qquad \;\;\;
[ Q_{D}, a_{\mu} (k) ] = - \varepsilon_{\mu\nu} k^{\nu} b(k),
\nonumber\\
&& \{ Q_{B}, c^{\dagger} (k) \} = 0,\;\;\;\;\; 
\qquad\;\;\;\;\;\;\;\;\; \{ Q_{D}, c^{\dagger} (k) \} = 
i \varepsilon^{\mu\nu} 
k_{\mu} a^{\dagger}_{\nu}, \nonumber\\
&& \{ Q_{B}, c (k) \} = 0, \;\;\;\;\;\;\qquad\;\;\;\;\;\;
\;\;\;\{ Q_{D}, c (k) \} = - i \varepsilon^{\mu\nu} k_{\mu} a_{\nu},
\nonumber\\
&& \{ Q_{B}, b^{\dagger} (k) \} = - i k^{\mu} a^{\dagger}_{\mu},
\qquad\;\;\;\;
\{ Q_{D}, b^{\dagger} (k) \} = 0, \nonumber\\
&& \{ Q_{B}, b(k) \} = + i k^{\mu} a_{\mu}, \qquad \;\;\;\;
\;\;\{ Q_{D}, b(k) \} = 0, 
\end{array} \eqno(5.5)
$$
where we have inserted the normal mode expansion (5.4) in the
symmetry transformation for the fields. Similary, the Casimir
operator $W$ generates the following commutation relations with
creation and annihilation operators:
$$
\begin{array}{lcl}
&& [ W , a^{\dagger}_{\mu} (k) ] = 
+ i k^2 \varepsilon_{\mu\nu} (a^{\nu})^{\dagger}, \qquad 
[ W,  a_{\mu} (k) ] = - i k^2 
\varepsilon_{\mu\nu} a^{\nu}, \nonumber\\
&& [ W, c (k) ]\; = \;[ W, c^{\dagger} (k) ]\; = \;[ W, b(k) ]\;
= \;[ W, b^{\dagger} (k) ] \;= 0.
\end{array} \eqno(5.6)
$$
It is clear that if we exploit the on-shell condition ($i.e.,
 k^2 = 0 $) for the fields, the commutation relations
generated by $W$ will be trivial.

Let us define the physical vacuum $|vac>$ of the theory as
$$
\begin{array}{lcl}
&&Q_{B}\;|vac>\;\; = \;\;Q_{D}\; |vac>\;\; =\; W\; |vac>\; =\;\; 0, \nonumber\\
&& a_{\mu} (k) \; |vac> \;=\; c (k)\; |vac> \;=\; b (k)\;|vac> = 0.
\end{array}\eqno(5.7)
$$
Now a single photon state with polarization $e_{\mu}$ can be created from
the physical vacuum by the application of a creation operator $
a^{\dagger}_{\mu}$; namely, $ e^{\mu}\; a^{\dagger}_{\mu}\;|vac> $.
We denote this state by $ | e,\; vac> $. Similary, a single photon
state with momentum $k_{\mu}$ can be represented as $ | k,\; vac>
= k^{\mu}\; a^{\dagger}_{\mu}\; |vac> $. Exploiting the commutation
relations of (5.5) and the physical state condition (5.7), this
state can be written as $ | k,\;vac> = Q_{B} (i b^{\dagger} 
(k) | vac >$). Thus, the normal gauge transformation
(with any arbitrary constant $\alpha$) can be expressed as

$$
\begin{array}{lcl}
| e + \alpha\; k, vac > = | e, vac > +  Q_{B} \bigl (i\;\alpha\;
b^{\dagger}(k)\bigr ) |vac >.
\end{array}\eqno(5.8) 
$$
According to the de Rham cohomology [9-12], all the  BRST exact states
are trivial. Thus, state $ | e + \alpha k, vac> $ is equivalent to
state $ |e, vac>$ which demonstrates the gauge invariance in the theory. 
Now let us concentrate on the physicality criteria
on one photon state with polarization $e_{\mu}$. Using the commutation
relations from (5.5), it is clear that
$$
\begin{array}{lcl}
Q_{B} | e,\; vac> = - (k \cdot e)\; c^{\dagger} (k)\; | vac > = 0,
\end{array}\eqno(5.9)
$$
which proves the transversality $ k\cdot e = 0$ of photon because
$ c^{\dagger} (k) |vac> $ is not a null state. For the
 langitudinal
or scalar photon for which $ k \cdot e \neq 0 $, we find that
$$
\begin{array}{lcl}
c^{\dagger}(k) | vac > = -\; \frac{1}{ k \cdot e}\; \;Q_{B}\; |e,\;vac>,
\end{array}\eqno(5.10)
$$
which is the statement of no-ghost theorem in the language of BRST 
cohomology. In fact, in the proof of unitarity of the non-Abelian gauge
theory, it is well known [22] that the ghost contributions cancel
the contributions coming from the longitudinal or scalar gluons. Hence,
ghosts are believed to exist in the virtual processes where there
is a gluon loop contribution. The meaning of equation (5.10) can
be stated in a different way. It says that any state with momentum
$k$, that is created by the application of $ c^{\dagger} (k)$ on the 
physical vacuum, is a BRST exact state if corresponding gauge photon 
of momentum $k$ is longitudinal or scalar. Thus, states corresponding to
the longitudinal or scalar photon are trivial states from the point of
view of BRST cohomology.

Now let us consider a dual state to the state 
$ | k, vac> $\footnote{ A dual state can be obtained from a state 
by the transformations $ k_{\mu} \rightarrow \pm i \varepsilon_{\mu\nu}
k ^{\nu}, C \rightarrow \pm i \bar C, \bar C \rightarrow
\pm i C$. This is the analogue of the Hodge $ {*} $ operation
of cohomology in the language of symmetry transformations. It
can be seen explicitly that $ Q_{B} |\psi> = - \;{*}\; Q_{D}
\;{*} |\psi> $, where $|\psi>$, in general, may depend on
$ k, C, \bar C $.}. This state can be written as: 
$\varepsilon_{\mu\nu} k^{\nu} (a^{\mu})^{\dagger} | vac > 
= i Q_{D} c^{\dagger} (k) |vac>$.
This shows that the dual state is a BRST co-exact state. Now
the physicality criterion on the one photon state $ |e,\; vac>$
w.r.t. $Q_{D}$ implies that $ \varepsilon_{\mu\nu} 
e^{\mu} k^{\nu} = 0$.  This is same as the transversality
condition on photon if we take into account the masslessness
 ($ k^2 = 0 $) condition. The precise expression is
$$
\begin{array}{lcl}
Q_{D}\; | e,\;vac> =\; \varepsilon_{\mu\nu}\; e^{\mu}\; k^{\nu}\; b^{\dagger}
(k)\;\; | vac > = 0.
\end{array}\eqno(5.11)
$$
If photons are not transverse then the above equation implies
the no-antighost theorem because the $ b^{\dagger} (k) |vac >$
state turns out to be BRST co-exact state. This, in turn, implies
that the $ b^{\dagger} (k) |vac>$ state is not a physical state as
far as the full BRST cohomology on the physical harmonic state
is concerned. Similarly, we can apply the Casimir operator $W$
on a single photon state $ |e, \;vac >$ and require the physicality
criterion which ultimately leads to the masslessness condition
$ k^2 = 0 $ of the photon. All these results are:
$$
\begin{array}{lcl}
Q_{B}\; |e,\;vac> &=& \;0\; \rightarrow\;\; \; k\cdot e = 0,\nonumber\\
Q_{D}\; |e,\;vac> &=& \;0\; \rightarrow
 \; \;\;
\varepsilon_{\mu\nu} e^\mu k^\nu = 0, \nonumber\\
W\;\; |e,\; vac> &=&\; 0\; \rightarrow\;\;  \;\;k^2 = 0.
\end{array}\eqno(5.12)
$$
For the free $U(1)$ gauge theory, the criteria of physicality
condition lead to the relations (5.12) which are consistent with 
one-another. In fact, the top two basic relations imply the third one.
Furthermore, the relations $ k \cdot e = 0 $ and
$ \varepsilon_{\mu\nu} e^{\mu} k^\nu = 0 $ of (5.12) respect
gauge invariance under transformations: $ e_{\mu} \rightarrow
e_{\mu} + \alpha \; k_{\mu}, e_\mu \; \rightarrow  e_\mu
+ \beta \varepsilon_{\mu\nu} k^\nu $ if $ k^2 = 0 $. Here 
$\alpha$ and $\beta$ are c-number constants. Normally, one defines
a harmonic (a single photon) state as the one which is annihilated
by the Laplacian operator ($ W | e, vac> = 0$). This condition, in turn,
implies $Q_{B} |e, vac> = 0, Q_{D} |e, vac> = 0$. It can be readily seen
that the condition $ k^2 = 0$ finds its solution in the
form of conditions $k \cdot e = 0 \;(i.e., k_{0} e_{0} = k e_{1})$ and
$ \varepsilon_{\mu\nu} e^{\mu} k^{\nu} = 0 \;(i.e., k_{0} e_{1} = k e_{0})$
which are the relations among the components of
the momentum vector $k_{\mu}$ and polarization vector $e_{\mu}$.
These relations emerge due to the conditions: $Q_{B} |e, vac> = 0,
Q_{D} |e, vac> = 0$ respectively. Now comparing the expression for
$e_{0}$ ($i.e.,  e_{0} = \frac{k e_{1}} {k_{0}} = \frac{k_{0} e_{1}}
{k}$), it can be seen that the masslessness condition emerges very
naturally. Thus, these relations on a single photon state do satisfy
the fact that $W|e, vac> = 0$  implies $Q_{B} |e, vac> = 0,
Q_{D} |e, vac> = 0$ in a subtle way.

It is obvious that the states 
$ c^\dagger(k)|vac>,\; b^{\dagger} (k) |vac> $ 
are BRST exact and co-exact respectively ($cf.$ (5.10) and (5.11)). 
Thus, these are not the physical states. It is interesting  to note that
a single photon state $ (a^{\dagger}_{\mu} (k) |vac>) $ in 2D
can also be written as the sum of BRST- and co-BRST exact states. This happens
here because of the topological nature of the theory. The existence of BRST-
and co-BRST symmetries enables one to decompose both the degrees of 
freedom of a single 2D photon into a component parallel to the momentum vector
(BRST exact state) and the other  component parallel to the polarization vector
(co-BRST exact state) (see, $e.g.$. eqns. (3.6), (5.8), (5.11)). Thus, a 2D 
photon is also not a physical state {\it per se}. This statement is
in conformity with its topological nature.\\

\noindent
{\bf  6 Topological Invariants}\\

\noindent
It is evident from equation (5.2) that for a single physical photon state,
we obtain conditions (5.12) due to the BRST cohomology. These are
mutually consistent with one-another. In other words, the validity
of any two of them implies the third condition. Thus, if
basic symmetries are the guiding principles,
the operation of $W$ on a single physical photon state is
superfluous because the symmetry generated by $W$ can be 
obtained from the ones generated by
$Q_{B}$ and $Q_{D}$. In fact, the presence of the basic BRST- 
and dual BRST symmetries are good enough to gauge away both the physical
degrees of freedom of photon in 2D. Further, as a consequence of the
presence of these two basic symmetries ($ \partial \cdot A = 0, 
\varepsilon_{\mu\nu} \partial^{\mu} A^{\nu} = 0$), the 2D physical
photon is forced to propagate on its mass-shell as well as on-shell 
($ \Box A_{\mu} = 0$).  Thus, free $U(1)$ gauge theory
becomes topological in nature [18]. In the framework of the BRST cohomology
and the Hodge decomposition theorem, this fact is encoded in the
vanishing of the Laplacian operator $W$
$$
\begin{array}{lcl}
W = {\displaystyle \int \; dx }\;
\frac{d} {dx} \;\bigl [ \; \frac{1}{2} (\partial \cdot A)^2 
- \;\frac{1}{2}\; E^2 \; \bigr ] \rightarrow 0 \qquad
\mbox{as} \qquad x \rightarrow \pm \infty,
\end{array}\eqno (6.1)
$$
when the equation of motion $ \partial_{\mu} E - \varepsilon_{\mu\nu}
\partial^\nu (\partial \cdot A) = 0 $ is exploited. Physically, the on-shell
expression for the Laplacian operator encompasses the physical degrees of
freedom left-over in the theory after some (or all) of them have been gauged
away by BRST- and co-BRST symmetries. Thus, expression (6.1) justifies
the topological nature of the 2D free $U(1)$ gauge theory. This
situation should be contrasted with the interacting $U(1)$ gauge theory
where the gauge field couples with the Dirac fields in 2D.
As it turns out, the on-shell expression of the Laplacian(Casimir) 
operator $W$ contains only the fermionic degrees of freedom 
(present in the theory) and it does not go to zero on the on-shell [23].

The topological nature of this theory is confirmed by the existence
of two sets of topological invariants w.r.t. conserved and on-shell
($ \Box C = \Box \bar C = 0$) nilpotent ($ Q_{B}^2 = 0, Q_{D}^2 = 0$)
BRST- and co-BRST charges. For the 2D compact manifold, these are
$$
\begin{array}{lcl}
I_{k} = {\displaystyle \oint}_{C_{k}} V_{k}, \qquad 
J_{k} = {\displaystyle \oint}_{C_{k}} W_{k},
\qquad (k = 0, 1, 2),
\end{array}\eqno(6.2)
$$
where $C_{k}$ are the k-dimensional homology cycles in the 2D manifold
and $V_{k}$ and $W_{k}$ are $k$-forms. For the free $U(1)$ gauge theory,
these forms are juxtaposed as
$$
\begin{array}{lcl}
&& V_{0} = - (\partial \cdot A)\; C, \;\;\;\;\;\;\qquad\;\;\;\;\;\;
\;\;\;\;\;\;\;\;\; W_{0} = E\; \bar C, \nonumber\\
&& V_{1} = \bigl [ i C \partial_{\mu} \bar C - A_{\mu} 
(\partial \cdot A) \bigr ]\; dx^\mu, \quad\;\;\;
W_{1} = \bigl [ \bar C \varepsilon_{\mu\nu} \partial^{\nu} C -
 i E \;A_{\mu} \bigr ]\; dx^\mu, \nonumber\\
&& V_{2} = i \bigl [ A_{\mu} \partial_{\nu} \bar C - \frac{\bar C} {2}
F_{\mu\nu} \bigr ]\; dx^\mu \wedge dx^\nu,\;\;
W_{2} = i \bigl [ \varepsilon_{\mu\rho} \partial^{\rho} C A_{\nu}
+ \frac{C}{2} \varepsilon_{\mu\nu} (\partial \cdot A) \bigr ]
dx^\mu \wedge dx^\nu.
\end{array}\eqno(6.3)
$$
It is starightforward to check that $V_{0}$ and $W_{0}$ are BRST-
and co-BRST invariants on the on-shell ($ \Box C = \Box {\bar C} = 0$)
because $ \delta_{B} (\partial \cdot A) = \eta\; \Box\; C $ and
$ \delta_{D} E = \eta\; \Box\; {\bar C} $. Thus, for all practical purposes,
it can be seen that one can take $ \delta_{B} (\partial \cdot A) = 0$
and $ \delta_{D} E = 0$ on the on-shell. The above BRST- and co-BRST
invariants ($cf.$ (6.2) and (6.3)) can be also written for the 
off-shell nilpotent BRST- and co-BRST transformations as the 
Lagrangian density (3.3) can be re-expressed as:
$$
\begin{array}{lcl}
{\cal L}_{\cal B} = {\cal B} E - \frac{1}{2}\;{\cal B}^2
+ B (\partial \cdot A)
+ \frac{1}{2}\; B^2 + i \bar C \Box C,
\end{array}\eqno(6.4)
$$
by introducing two auxiliary fields ${\cal B}$ and $B$. The off-shell nilpotent
BRST- and co-BRST symmetry transformations for the above Lagrangian density
(6.4) are same as the transformations (3.2) except for  the ghost- and
auxiliary  fields. These additional transformations are:
$$
\begin{array}{lcl}
\delta_{B} \bar C &=& i \eta B, \qquad \;
\delta_{B} B = 0, \qquad \;\delta_{B} {\cal B} = 0, \nonumber\\
\delta_{D} C &=& - i \eta {\cal B}, \qquad 
\delta_{D} {\cal B} = 0, \qquad \delta_{D} B = 0.
\end{array}\eqno(6.5)
$$
Now the BRST- and co-BRST invariants (w.r.t. off-shell nilpotent
BRST- and co-BRST transformations) can be expressed in terms
of the auxiliary fields ${\cal B}$ and $B$ by using the equations of motion
$ {\cal B} = E$ and $ B = - (\partial \cdot A)$
($cf.$ (6.2) and (6.3)).  Furthermore, it can be 
checked that $V_{2}$ and $W_{2}$ are closed ($ d V_{2} = 0$) and co-closed
($ \delta W_{2} = 0$) respectively. The ghost number for $V_{k}$ and $W_{k}$
are ($ +1, 0, -1$) and ($ -1, 0, +1$) respectively as
can be seen from the following commutation relations:
$$
\begin{array}{lcl}
&& i [ Q_{g}, V_{k} ] = (-1)^{1 - k}\; (k - 1) V_{k}, \nonumber\\
&& i [ Q_{g}, W_{k} ] = (-1)^{1 - k}\; (1 - k) W_{k},
\end{array}\eqno(6.6)
$$
where $ k = 0, 1, 2$ and $Q_{g}$ is the ghost charge. The above BRST-
and co-BRST invariants obey the following relations (see, $e.g.$, [24,25])
$$
\begin{array}{lcl}
\delta_{B} V_{k} &=& \eta\; d\; V_{k-1}, \qquad \;\;\;\;
d = dx^\mu\; \partial_{\mu}, \nonumber\\
\delta_{D} W_{k} &=& \eta \delta\; W_{k-1}, \;\qquad \;\;\;
\delta = i dx^\mu\; \varepsilon_{\mu\nu} \partial^\nu,
\end{array}\eqno(6.7)
$$
where, as the above expression shows, $d$ and $\delta$ are exterior-
and dual exterior derivatives respectively on the 2D compact manifold.
Both these sets of topological invariants are connected with each-other
by the duality transformations: $ (\partial \cdot A) \rightarrow
i E, C \rightarrow i \bar C, \partial_{\mu} \rightarrow i 
\varepsilon_{\mu\nu} \partial^\nu (B \rightarrow - i {\cal B})$ 
as $ I_{k} \rightarrow J_{k}$. Thus,
these invariants are not entirely independent of each-other
as, in some sense, the exterior derivative $d$ and the dual exterior
derivative $\delta (= \pm * d *)$ are not completely independent.

It is interesting to verify that the Lagrangian density (3.3), modulo
some total derivatives, can be written as the sum of anticommutators with
BRST- and co-BRST charges:
$$
\begin{array}{lcl}
{\cal L}_{B} = \{ Q_{D}, S_{1} \} + \{ Q_{B}, S_{2} \},
\end{array}\eqno(6.8)
$$
where $S_{1}= \frac{1}{2} E C, S_{2} = - \frac{1}{2} (\partial \cdot A)
\bar C$. Using the fact that $Q_{r} (r = B, D)$ is the generator of the
transformations $ \delta_{r} \Phi = - i \eta [ \Phi, Q_{r} ]_{\pm}$,
where $(+)-$ stands for the (anti)commutator depending on the generic
field $\Phi$ being (fermionic)bosonic, it can be checked that:
$ \eta {\cal L}_{B} = \frac{1}{2} \; \delta_{D} [ i E C] - \frac{1}{2}\;
\delta_{B} [ (\partial \cdot A) \bar C ]$. 
Mathematically, this observation shows that
the 2D free $U(1)$ gauge theory is similar in 
{\it outlook}  as the Witten type
topological field theories [25] but quite different from the Schwarz type 
topological theories [26]. To be very precise, there is a bit of difference
with the Witten type theories as well. This is primarily because of the fact
that in our discussion there are two nilpotent charges w.r.t. which
topological invariants and the BRST cohomology are defined whereas in 
the Witten type theories there is only one nilpotent BRST charge [25]
which is obtained by combining a topological shift symmetry with the local
gauge symmetry. It is clear, however, that there are no shift symmetries
in our whole discussions. Thus, from 
{\it symmetry point of view}, the free 2D
$U(1)$ gauge theory is more like Schwarz type topological theories where only
local gauge symmetries exist. 
As a consequence, this 2D free theory is a new type of topological
field theory which captures together some of the
salient features of both Witten- and Schwarz type topological theories.
For both ($i.e.$, Witten as well as Schwarz) type
of theories the energy-momentum tensor ($T_{\alpha\beta}$) is always
a BRST (anti)commutator. It can be seen that for the free 2D $U(1)$ gauge
theory, the expression for the {\it symmetric} $T_{\alpha\beta}$ is:
$$
\begin{array}{lcl}
T_{\alpha\beta} &=& - \frac{1}{2} \bigl [\; \varepsilon_{\alpha \rho}\;E
+ \eta_{\alpha \rho} (\partial \cdot A) \;\bigr ]\; \partial_{\beta} A^{\rho}
- \frac{1}{2} \bigl [\; \varepsilon_{\beta\rho} \; E + \eta_{\beta\rho}
(\partial \cdot A) \;\bigr ]\; \partial_{\alpha} A^{\rho}\nonumber\\
&-& i\; \partial_{\alpha} \bar C \;\partial_{\beta} C - i\; \partial_{\beta} 
\bar C \;\partial_{\alpha} C - \eta_{\alpha\beta}\; {\cal L}_{B},
\end{array}\eqno(6.9)
$$
where ${\cal L}_{B}$ is the Lagrangian density of equation (3.3) (or (6.8)).
This energy momentum tensor, modulo some total
derivatives, can be re-expressed as:
$$
\begin{array}{lcl}
\eta \;T_{\alpha\beta} &=&  \frac{i}{2} \delta_{B} \bigl [\; 
 \partial_{\alpha} \bar C\; A_{\beta} +  \partial_{\beta} \bar C\; A_{\alpha}
+  \eta_{\alpha \beta} (\partial \cdot A)\; \bar C \;\bigr ] \nonumber\\
&+& \frac{i}{2}
\delta_{D} \bigl [\; \partial_{\alpha} C \; \varepsilon_{\beta \rho} A^{\rho}
+ \partial_{\beta} C\; \varepsilon_{\alpha \rho} A^{\rho} - \eta_{\alpha\beta}
E\; C\;\bigr ],
\end{array}\eqno(6.10)
$$
where $\delta_{B}$ and $\delta_{D}$ correspond to transformations in (3.2). 
In terms of
BRST- and dual BRST charges, we can write $T_{\alpha\beta}$ as:
$$
\begin{array}{lcl}
T_{\alpha\beta} =  \{ Q_{B}, V_{\alpha\beta}^{(1)} \}
+ \{ Q_{D}, V_{\alpha\beta}^{(2)} \},
\end{array}\eqno(6.11)
$$
where
$$
\begin{array}{lcl}
V_{\alpha\beta}^{(1)} &=& \frac{1}{2} \bigl [ \;
 \partial_{\alpha} \bar C A_{\beta} +  \partial_{\beta} \bar C A_{\alpha}
+  \eta_{\alpha \beta} (\partial \cdot A) \bar C \;\bigr ],
\nonumber\\ 
V_{\alpha \beta}^{(2)} &=& \frac{1}{2}
 \bigl [\; \partial_{\alpha} C \; \varepsilon_{\beta \rho} A^{\rho}
+ \partial_{\beta} C \;\varepsilon_{\alpha \rho} A^{\rho} - \eta_{\alpha\beta}
E\; C \; \bigr ].
\end{array}\eqno(6.12)
$$
Thus, for the theory under discussion, the form of
energy-momentum tensor is just like
Witten as well as Schwarz type of topological theories.
It is now straightforward to argue that the partition functions as well
as the expectation values of the BRST invariants, co-BRST invariants
and the topological invariants are metric independent. The key point
to show this fact in the framework of the BRST cohomology and the
Hodge decomposition theorem is the requirement that $Q_{B} |phys> = 0,
Q_{D} |phys> = 0$ (see, $e.g.$, Ref. [18] for details) and the metric
independence of the path integral measure (see, $e.g.$, Ref [24]
for details). We have
taken here only the flat Minkowski metric. However, our arguments and
discussions are valid even if we take a nontrivial metric. In Ref. [24],
it has been argued that the fermionic-bosonic symmetry of the BRST formalism
is good enough to show that the path integral measure will be independent
of the choice of the metric.\\

\noindent
{\bf  7 Discussion}\\

\noindent
It is clear  that the usual nilpotent BRST transformations
correspond to a symmetry in which the two-form $ F = d A $
($e.g.$, electric field $E$ in 2 D) of the U(1) gauge theory remains invariant.
The nilpotent dual-BRST charge is the generator of a transformation in which
the gauge-fixing term $((\partial \cdot A)= \delta  A)$ remains invariant.
The anticommutator of these two transformations corresponds to a symmetry that 
is generated by the Casimir operator for the whole algebra. Under this 
conserved operator, it is the ghost fields that remain invariant. We see from
algebra (4.3) and the ghost number considerations in (4.4) that the
generators $ Q_{B}, Q_{D}$ and $ W $ of symmetry transformations 
correspond to the
geometrical quantities $ d, \delta$ and $ \Delta $ of  differential geometry
which describe  the de Rham 
cohomology of differential forms on a compact manifold. 
It is, however, the peculiarity of the BRST formalism that these geometrical
quantities can be
identified with two conserved charges. For instance, in addition to the previous
identifications, $ d $ and $\delta $ could also be identified with $Q_{AB}$
and $ Q_{AD}$. Thus, $W$ and $\Delta$ can be expressed in two different ways :
$ W = \{ Q_{B}, Q_{D} \} = \{ Q_{AB}, Q_{AD} \} $ and 
$ \Delta = d \delta + \delta d = \bar d \bar \delta + \bar \delta \bar d $.
This shows that the compact manifold, on which $d, \delta$ and $\Delta$ 
are defined, should be a complex manifold as far
as the analogy with BRST cohomology of physical states in the quantum
Hilbert space is concerned.

We know that the classical  ($e.g.$, electric and magnetic) fields correspond
to the two-form $ F = d A $ which turn out to be gauge- and 
BRST invariant because
of the structure of the field strength tensor $ F_{\mu\nu} = \partial_{\mu} 
A_{\nu} - \partial_{\nu} A_{\mu} $ and the specific tranformation on $A_{\mu}$.
We know from the celebrated Aharonov-Bohm effect that it is the vector potential
($i.e.$ one-form $ A = A_{\mu} dx^{\mu}$) that plays 
decisive role at the quantum
level and not the classical electric or magnetic fields $ F = d A $.  Thus, 
symmetry
transformation that leaves the gauge-fixing term $( \delta \;A )$ invariant is a
quantum mechanical symmetry (right from the outset) by its very nature. It is 
precisely 
due to this reason that the physical state condition with dual-BRST charge
$(Q_{D}\;|phys> = 0)$ leads to a quantum mechanical restriction on the physical
state
when $U(1)$ gauge field is coupled to the Dirac fields. In fact, it has been 
shown in Ref. [23] that the dual-BRST transformation $ \delta_{D} A_{\mu} 
= - \eta \varepsilon_{\mu\nu} \partial^{\nu} \bar C $ corresponds to the 
chiral transformation on the Dirac fields for fermions in 2D.
It is obvious that the  condition $ Q_{D} \;|phys> = 0 $ 
would shed some light on the 2D anomaly
term $(E \sim \varepsilon_{\mu\nu}\; F^{\mu\nu})$ in QED. Thus, 
full strength of the 
BRST cohomology might provide a clue to the well known result 
that in 2D, the ``anomalous'' gauge theory is consistent, unitary 
and amenable to particle interpretation [27,20]. 
To begin with, the theory under discussion (Ref. [23]) is not a chiral gauge
theory in 2D. The chiral symmetry in our discussion is respected at
the quantum level as it corresponds to the dual BRST symmetry.

For the free $U(1)$ gauge theory,
all the conditions in equation (5.12) are consistent with one-another
and physically they imply the masslessness and transversality
of photon. In fact, the existence of gauge symmetries: $ e_{\mu}
\rightarrow  e_{\mu} + \alpha \; k_{\mu}, e_{\mu} \rightarrow
e_{\mu} + \beta\; \varepsilon_{\mu\nu} \; k^{\nu} $
(for $\alpha$ and $\beta$ being arbitrary constants), imply that
both the degrees of freedom of photon $A_{\mu}$ can be gauged
away in two dimensions of spacetime by symmetry considerations
alone and masslessness condition ($k^2 = 0$) emerges due to these 
symmetries.  Thus, this theory becomes topological in nature 
as there are no propagating degrees of freedom left in the 
theory [18].
Normally, it is the masslessness $ k^2 = 0$  and transversality 
$ k \cdot e = 0 $ criteria that are sufficient to get rid of
both the degrees of freedom of photon in 2D. The former comes out
from the equation of motion ($\Box A_{\mu} = 0$) and the latter is
a gauge-fixing constraint imposed on the theory ($\partial \cdot A = 0$).
In our discussion, however,
the transversality of the photon ($ k \cdot e = 0$) and
the relation ($ \varepsilon_{\mu\nu} e^\mu k^\nu = 0$) between 
the polarization vector $e_{\mu}$ and momentum vector $k_{\mu}$,
emerge because of the presence of BRST- and co-BRST symmetries.
In fact, these two  basic (BRST and co-BRST) symmetries gauge away both
the physical degrees of freedom of photon and, in a subtle way, these are
the solutions to the 
masslessness  condition $ k^2 = 0$ which emerges
due to $W |e, vac> = 0$ (see, $e.g.$, eqn. (5.12)). However, 
it due to  the topological 
nature of this theory  that $W \rightarrow 0$ when equations of
motion are exploited.
The existence of the
topological invariants on 2D compact manifold confirms the topological
nature of this theory in a cogent way. Such arguments have also been
provided for the proof of topological nature of 2D free non-Abelian 
gauge theory (having no interaction with matter fields) [28]. It will
be interesting to understand the BRST cohomology and Hodge decomposition
theorem for the interacting theory in two- and four dimensions of spacetime
where there is a coupling between the (non)Abelian gauge field and  
matter fields [29]. Some of these results 
have been obtained for the interacting $U(1)$ gauge theory in Ref. [23]
where the dual BRST transformation on the gauge field 
has been shown to correspond to the
chiral transformation on the Dirac fields in 2D spacetime.

The upshot of our whole discussion is to capitalize
on the insight gained  in the case of 2D free $U(1)$ gauge theory
and generalize these ideas to 4D. In fact, it is already known that
the duality of the Maxwell equations and the chirality of the massless 
fermions are very intimately connected even in 4D (see, $e.g.$, Ref. [30]). 
It will be interesting to study the implication of the dual BRST symmetry
in four dimensional spacetime when two-potentials are present in the field
strength tensor $ F_{\mu\nu} = \partial_{\mu} A_{\nu} -
\partial_{\nu} A_{\mu} +  \varepsilon_{\mu\nu\lambda\xi}\; \partial^{\lambda}
\;V^{\xi} $ where $V_{\mu}$ is an axial-vector [30]. This understanding
might shed some light on the axial-vector anomaly in 4D and the Dirac
quantization condition in QED. The latter is connected to the duality in 
electric-magnetic couplings due to presence of two potentials and
their coupling with matter (Dirac) fields. In the case
of 4D non-Abelian gauge theories, the complete understanding of 
$ Q_{B}, Q_{D}$ and $ W$ might provide some insight into the problem of 
confinement of quarks and gluons in the language of BRST formalism 
[31,32]. These are some of the issues for future investigations [29].\\

\noindent
{\bf Aknowledgements}\\

\noindent
Part of this work was done at The Abdus Salam ICTP, Trieste (Italy) and
JINR, Dubna (Moscow).  Fruitful conversations with the members of  
theoretical high energy physics groups at ICTP and JINR are gratefully 
acknowledged. It is a pleasure to thank the referee for his 
very useful and enlightening comments.\\ 

\baselineskip = 12pt

\noindent
{\bf References}\\

\begin{enumerate} 
\item  C. Becchi, A. Rouet and R. Stora, {\it Commun. Math. Phys.}
       {\bf 42}, 127 (1975);  {\it Ann. Phys. (N.Y.)} {\bf 98}, 287 (1976).
\item  I. V. Tyutin, {\it Lebedev Preprint} FIAN report no {\bf 39}, 
       (1975) (unpublished).
\item  K. Sundermeyer, {\it Constrained Dynamics: Lecture Notes
       in Physics} Vol. {\bf 169}, (Springer-Verlag, Berlin, New York, 1982).
\item  K. Nishijima, in {\it  Progress in Quantum Field Theory} eds.
       H. Ezawa and S. Kamefuchi, (North- Holland, Amsterdam, 1986) p.99.
\item  M. Henneaux and C. Teitelboim,  
         {\it  Quantization of Gauge Systems} 
       (Princeton University Press, Princeton, 1992).
\item  N. Nakanishi and I. Ojima, {\it Covariant Operator Formalism of
       Gauge Theories and Quantum Gravity}, (World Scientific, 
       Singapore, 1990).
\item  I. A. Batalin and I. V. Tyutin, 
       {\it Int. J. Mod. Phys.} {\bf A6}, 3255 (1991).
\item  I. A. Batalin, S. L. Lyakhovich and I. V. Tyutin, 
       {\it Mod. Phys. Lett.}
       {\bf A7}, 1931 (1992); {\it Int. J. Mod. Phys.} {\bf A10}, 1917 (1995).
\item  T. Eguchi, P. B. Gilkey  and A. J. Hanson,  
       {\it Phys. Rep.} {\bf 66}, 213 (1980). 
\item  J. W. van Holten,  {\it Phys. Rev. Lett.} {\bf 64},
        2863 (1990); {\it Nucl. Phys.} {\bf B339}, 158 (1990).
\item  S. Mukhi  and N. Mukunda,  {\it Introduction  to Topology,
       Differential Geometry and Group Theory for Physicists}, (Wiley Eastern
       Ltd., New Delhi, 1990). 
\item  H. Aratyn,  {\it J. Math. Phys.}  {\bf 31}, 1240 (1990).
\item  D. McMullan and M. Lavelle, {\it Phys. Rev. Lett.} {\bf 71},
       3758 (1993);
       {\it ibid.} {\bf 75}, 4151 (1995).
\item  V. O. Rivelles,  {\it Phys. Rev. Lett.} {\bf 75}, 4150
       (1995); {\it Phys. Rev.} {\bf D 53}, 3257 (1996).
\item  H. S. Yang  and B. -H. Lee,    {\it J. Math. Phys.}
       {\bf 37}, 6106 (1996).
\item  R. Marnelius,   {\it Nucl. Phys.} {\bf B494}, 346 (1997).
\item  T. Zhong and D. Finkelstein,   {\it Phys. Rev. Lett.}
       {\bf 73}, 3055 (1994);  {\it ibid.} {\bf 75}, 4152 (1995).
\item  D. Birmingham, M. Blau, M. Rakowski and G. Thompson, {\it Phys. Rep.}
       {\bf 209}, 129 (1991).
\item  R. P. Malik, {\it Int. J. Mod. Phys.} {\bf A12}, 231 (1997),\\
       see also, {\it Commuting conserved quantities in nonlinear realizations
       of $W_{3}$}, JINR- report no. {\bf E2-96-120} (1996).
\item  R. P. Malik,  {\it Phys. Lett.} {\bf 212B}, 445 (1988).
\item  S. Weinberg,  
        {\it The Quantum Theory of Fields: Modern Applications }
       {\bf V.2} (Cambridge University Press, Cambridge, 1996).
\item  I. J. R. Aitchison and A. J. G. Hey,  
       {\it Gauge Theories in Particle
       Physics: A Practical Introduction} (Adam Hilger, Bristol, 1982).
\item  R. P. Malik,  {\it Dual BRST Symmetry in QED} : 
       hep-th/ 9711056.
\item  R. K. Kaul and R. Rajaraman, {\it Phys. Lett.} {\bf B265}, 335 (1991);
       {\it ibid.} {\bf B249}, 433 (1990).
\item  E. Witten, {\it Commun. Math. Phys.} {\bf 121}, 351 (1988).
\item  A. S. Schwarz, {\it Lett. Math. Phys.} {\bf 2}, 217 (1978).   
\item  R. Jackiw and R. Rajaraman,  {\it Phys. Rev. Lett.}  
       {\bf 54}, 1219 (1985).
\item  R. P. Malik, {\it Mod. Phys. Lett.} {\bf A14}, 1937 (1999).
\item  R. P. Malik, in preparation.
\item  R. P. Malik and T. Pradhan,   {\it Z. Phys.}
           {\bf C 28}, 525 (1985).
\item  K. Nishijima,   {\it Int. J. Mod. Phys.} {\bf A9}, 3799 (1994) ; 
       {\it ibid.} {\bf A10}, 3155 (1995).
\item  T. Kugo and I. Ojima,  {\it Prog. Theor. Phys. (Suppl)} 
       {\bf 66}, 1 (1979);   {\it Phys. Lett.} {\bf 73B}, 459 (1978).
\end{enumerate}
\end{document}